\documentclass[aps,prd,reprint,superscriptaddress,nofootinbib,amsmath,amssymb,floatfix]{revtex4-2}
\usepackage{graphicx}
\usepackage{bm}
\usepackage{braket}
\usepackage{mathrsfs}
\usepackage{dcolumn}
\usepackage{booktabs}

\begin{document}
	\title{Unified Exact Cylindrical Dirac Modes and Symmetry-Resolved Quantum Geometry}
	
	\author{Zhongze Guo}
	\email{guozhongze007@gmail.com}
	\affiliation{Department of Physics and Institute of Theoretical Physics, University of Science and Technology Beijing, Beijing 100083, China}
	
	\author{Bei Xu}
	\email{xubei0903@163.com}
	\affiliation{Institute for Advanced Study, Tsinghua University, Beijing 100084, China}	
	
	\author{Qiang Gu}
	\email[Corresponding author: ] {qgu@ustb.edu.cn}
	\affiliation{Department of Physics and Institute of Theoretical Physics, University of Science and Technology Beijing, Beijing 100083, China}

	\begin{abstract}
Exact cylindrical solutions of the free Dirac equation provide natural single-particle modes for a broad class of axially symmetric relativistic fermion systems, including electron vortices, twisted-particle scattering, rotating matter, and cylindrical field quantization, but are commonly formulated in different internal bases. We derive the general regular positive-energy cylindrical solution at fixed energy, transverse and longitudinal momenta, and total angular momentum, and show how the commonly used spin-polarized, separation, helicity, and transverse-helicity modes are embedded in the resulting two-dimensional solution space. A conserved transverse operator resolves the residual doublet into two symmetry-defined branches. Their parameter-dependent eigenspaces admit a gauge-invariant quantum-geometric characterization, with opposite Berry curvatures, a common quantum metric, and saturation of the two-level metric--curvature relation. We further derive symmetry constraints on branch conversion and the corresponding reduced two-state dynamics. The resulting framework connects mode classification, symmetry resolution, quantum geometry, and dynamics across distinct cylindrical Dirac settings.
\end{abstract}

\maketitle

\section{Introduction}

Exact cylindrical solutions of the Dirac equation provide natural single-particle modes in relativistic systems with a distinguished symmetry axis. They describe relativistic electron vortex beams, which have been generated experimentally as carriers of orbital angular momentum~\cite{Masaya10,Verbeeck10,McMorran11}, and underlie their relativistic angular-momentum and spin-orbit structure~\cite{Bliokh07,Bliokh2011,Bliokh17pra,Bliokh2017Review,electron17rmp,Birula17,Barnett17,Karlovets18pra}. Closely related helicity-adapted twisted states are widely used in non-plane-wave scattering theory~\cite{Ivanov2011PRD,IvanovSerbo2011,Serbo2015TwistedScattering,Karlovets2017JHEP,Karlovets2017PRA,Ivanov23,Zou2023Innovation}, while other cylindrical bases appear in longitudinally boost-invariant field quantization~\cite{Bjorken1983,Mihaila2006,Mihaila2009PRD}. Transverse-helicity cylindrical states, introduced in earlier cylindrical quantization work, have also been employed in rotating relativistic matter and confined Dirac systems~\cite{BalantekinDeWeerd1995,JiangLiao2016,Khosravi2019PRB}. Although these settings describe physically distinct phenomena, they share the same free cylindrical Dirac sector at the single-particle level. This sector therefore serves as a recurring kinematic building block rather than a mode construction tied to a single application.

The diversity of applications is accompanied by a diversity of internal bases. Different physical problems naturally select different resolutions of the same two-dimensional positive-energy cylindrical sector: spin-polarized modes are standard in relativistic vortex-beam descriptions, helicity-adapted states are convenient when scattering amplitudes are assembled from plane-wave components on a momentum cone, the Mihaila--Dawson--Cooper modes arise from a transverse separation used in boost-invariant field quantization, and transverse-helicity states are particularly natural in rotating and confined Dirac problems~\cite{BalantekinDeWeerd1995,JiangLiao2016,Khosravi2019PRB}. The spin-polarized and helicity descriptions are related by a momentum-dependent rotation associated with the massive-particle little group~\cite{Wigner1939}. These constructions are well adapted to their respective applications, but they do not by themselves expose the common regular solution space from which the different bases arise. The present analysis constructs this common space explicitly and identifies the symmetry resolution that organizes it.

The separation of a first-order Dirac system in curvilinear coordinates is more involved than that of the corresponding scalar equation~\cite{Villalba1989,Villalba1990}. The present analysis addresses two related questions at the level of the four-component free Dirac equation: the construction of the general regular cylindrical solution at fixed $(E,\kappa,k_z,J_z)$, and the identification of a conserved operator that resolves the remaining two-dimensional freedom within the transverse cylindrical sector.

A generalized series expansion, introduced without selecting an internal basis, yields a regular family containing a single free complex parameter $\lambda$. The conventional spin-polarized vortex pair and the cylindrical separation modes of Mihaila, Dawson, and Cooper are obtained as special members of this family, while helicity-adapted twisted states follow by unitary recombination within the same two-dimensional solution space. The construction therefore places exact cylindrical modes introduced in different representations and physical contexts within a common solution family.

The residual twofold freedom is naturally resolved by the conserved transverse operator
\begin{equation}
K=\beta(\boldsymbol{\Sigma}\times\mathbf p)_z .
\label{eq:K-intro}
\end{equation}
It obeys
$[K,H]=[K,p_z]=[K,J_z]=0$
and $K^2=p_\perp^2$. For $\kappa>0$, its eigenvalues $\eta\kappa$, $\eta=\pm1$, define two exact cylindrical branches. The overall sign of $K$ is conventional and only interchanges the branch labels; Eq.~\eqref{eq:K-intro} fixes the convention used throughout this paper. This operator supplies a symmetry-adapted resolution of the same regular solution space that is represented in conventional spin and helicity bases. Up to normalization and phase/sign conventions, its eigenstates coincide with the transverse-helicity cylindrical spinors of earlier work~\cite{BalantekinDeWeerd1995,JiangLiao2016,Khosravi2019PRB}. The contribution of the present construction is to derive the general regular family before choosing such a resolution and then to place the transverse-helicity, spin-polarized, helicity, and separation bases within that common family. This organization also provides the setting for a symmetry-resolved analysis of their quantum geometry and perturbative dynamics.

The quantum-geometric characterization follows directly from this symmetry resolution rather than from an additional choice of basis. At each $(\kappa,k_z)$, the free positive-energy Dirac sector is two dimensional; the parent degenerate sector therefore carries the non-Abelian Berry structure familiar from massive Dirac theory~\cite{ShankarMathur1994,ChenPangPuWang2014}. The conserved operator $K$ resolves this doublet into two one-dimensional eigenspaces. When $(\kappa,k_z)$ varies, the resulting eigenspaces themselves vary in Hilbert space even though the two branches share the same free dispersion. Their joint spectral projectors provide a basis- and phase-independent description of this variation, and the corresponding quantum geometric tensors~\cite{ProvostVallee1980,Berry1984,Simon1983,Xiao2010,Kolodrubetz2017,GrafPiechon2021} quantify its symmetric and antisymmetric components through the quantum metric and Berry curvature. In this sense, the geometry is a property of the symmetry-selected branch subspaces, not an Abelian structure assigned to arbitrary basis vectors in the unresolved doublet. The two branches are found to have opposite Berry curvature and the same quantum metric, with saturation of the two-level metric--curvature relation. This construction concerns the internal Dirac sector and does not include the additional parameter dependence of a spatially regularized Bessel envelope.

The dynamical implications of the symmetry are obtained from branch-resolved selection rules and from the decomposition of a general perturbation into $K$-even and $K$-odd components. For a scalar potential, the commutator $[K,V]$ implies exact branch conservation for a longitudinal profile $V(z,t)$ and branch-off-diagonal coupling in the presence of a transverse radial gradient. Projection onto an isolated positive-energy doublet gives a closed two-state Hamiltonian whose off-diagonal matrix elements determine the conversion probability. The Hadamard-form relation between the conventional spin-polarized basis and the $K$ basis further permits branch-dependent amplitudes and phases to be inferred from conventional-mode populations, without assigning an energy splitting to the degenerate free branches.

The paper is organized as follows. Section~II derives the general regular cylindrical family and relates it to standard exact modes. Section~III derives the transverse conserved operator and its branch eigenstates. Section~IV compares the branch basis with conventional spin and helicity descriptions. Section~V develops the intrinsic quantum geometry of the $K$ branches. Section~VI discusses symmetry protection, transverse-gradient-induced branch conversion, the reduced two-state dynamics, and possible procedures for coherent branch analysis. Section~VII summarizes the results.
    \section{Unified exact cylindrical Dirac family}
    \label{sec:generalized-series}
    
    This section constructs the general regular free cylindrical family at fixed $(E,\kappa,k_z,J_z)$ without selecting an internal basis in advance. Several exact cylindrical modes used in the literature are then identified as particular representatives of the resulting two-dimensional solution space. The transverse branch-resolving symmetry is derived independently in Sec.~\ref{sec:K-operator}.
    
    \subsection{Generalized series solution of the cylindrical Dirac equation}
    \label{subsec:generalized-series-solution}
    
    The relativistic electron is described by the Dirac equation
    \begin{equation}
    	\hat{H}|\psi\rangle=
    	\left(-ic\hbar\,\vec{\alpha}\cdot\vec{\nabla}+\beta mc^{2}\right)|\psi\rangle
    	=
    	E|\psi\rangle,
    	\label{eq:ham}
    \end{equation}
    where $m$ is the electron rest mass, $E$ is the eigenenergy, and
    \(
    |\psi\rangle=(\psi_1,\psi_2,\psi_3,\psi_4)^T
    \)
    is a four-component spinor. In the standard Dirac-Pauli representation,
    \[
    \beta=
    \begin{pmatrix}
    	I & 0\\
    	0 & -I
    \end{pmatrix},
    \qquad
    \vec{\alpha}=
    \begin{pmatrix}
    	0 & \vec{\sigma}\\
    	\vec{\sigma} & 0
    \end{pmatrix},
    \]
    with $\vec{\sigma}=(\sigma^x,\sigma^y,\sigma^z)$ the Pauli matrices and $I$ the $2\times2$ identity matrix.
    
    To describe cylindrical vortex states, it is convenient to rewrite the Dirac equation in cylindrical coordinates $(r,\theta,z)$. A useful way to do this is through the complex cylindrical basis
    \begin{equation}
    	\vec{\nabla}
    	=
    	\sum_{\mu}(-1)^\mu \vec e_\mu \nabla_{-\mu}
    	=
    	-\vec e_{+1}\nabla_{-1}
    	+\vec e_0\nabla_0
    	-\vec e_{-1}\nabla_{+1},
    	\label{eq:nabla-expansion}
    \end{equation}
    where
    \begin{align}
    	\nabla_{+1}
    	&=
    	-\frac{1}{\sqrt2}
    	\left(\frac{\partial}{\partial x}+i\frac{\partial}{\partial y}\right)
    	=
    	-\frac{e^{i\theta}}{\sqrt2}
    	\left(\frac{\partial}{\partial r}
    	+\frac{i}{r}\frac{\partial}{\partial\theta}\right),\\
    	\nabla_{-1}
    	&=
    	\frac{1}{\sqrt2}
    	\left(\frac{\partial}{\partial x}-i\frac{\partial}{\partial y}\right)
    	=
    	\frac{e^{-i\theta}}{\sqrt2}
    	\left(\frac{\partial}{\partial r}
    	-\frac{i}{r}\frac{\partial}{\partial\theta}\right),\\
    	\nabla_0&=\frac{\partial}{\partial z}.
    \end{align}
    With this notation, the free Dirac equation becomes the coupled cylindrical system
    \begin{align}
    	\frac{E-mc^2}{ic}\psi_1
    	+e^{-i\theta}\left(\hbar\frac{\partial}{\partial r}
    	-\frac{i\hbar}{r}\frac{\partial}{\partial\theta}\right)\psi_4
    	+\hbar\frac{\partial}{\partial z}\psi_3
    	&=0, \nonumber\\
    	\frac{E-mc^2}{ic}\psi_2
    	+e^{i\theta}\left(\hbar\frac{\partial}{\partial r}
    	+\frac{i\hbar}{r}\frac{\partial}{\partial\theta}\right)\psi_3
    	-\hbar\frac{\partial}{\partial z}\psi_4
    	&=0, \nonumber\\
    	\frac{E+mc^2}{ic}\psi_3
    	+e^{-i\theta}\left(\hbar\frac{\partial}{\partial r}
    	-\frac{i\hbar}{r}\frac{\partial}{\partial\theta}\right)\psi_2
    	+\hbar\frac{\partial}{\partial z}\psi_1
    	&=0, \nonumber\\
    	\frac{E+mc^2}{ic}\psi_4
    	+e^{i\theta}\left(\hbar\frac{\partial}{\partial r}
    	+\frac{i\hbar}{r}\frac{\partial}{\partial\theta}\right)\psi_1
    	-\hbar\frac{\partial}{\partial z}\psi_2
    	&=0.
    	\label{eq:cylindrical-dirac-system}
    \end{align}
    
    Previous treatments frequently reduce the first-order spinor system to second-order equations by additional transformations or in the presence of external fields. Here the free cylindrical Dirac equation is solved in its original first-order form, retaining the full spinor structure throughout the derivation.
    
    We therefore employ a generalized series-expansion ansatz. Each spinor component is written as
    \begin{equation}
    	\psi_s(r,\theta,z)=R_s(r)\,e^{in_s\theta}\,e^{ip_z z/\hbar}.
    \end{equation}
    Consistency of the angular factors in Eq.~\eqref{eq:cylindrical-dirac-system} fixes
    \begin{equation}
    	n_1=n_3=n,\qquad n_2=n_4=n+1,
    \end{equation}
    so that all four components have the common total-angular-momentum eigenvalue $J_z=(n+\tfrac12)\hbar$. The radial functions are expanded as
    \begin{equation}
    	R_s(r)=r^\alpha\sum_{k=0}^{\infty} C_k^s r^k.
    	\label{eq:radial-series}
    \end{equation}
    Substituting this ansatz into Eq.~\eqref{eq:cylindrical-dirac-system}, and requiring that the coefficient of each power of $r$ vanish, yields four coupled recurrence relations. With
    \begin{equation}
    	\varepsilon_\pm\equiv \frac{E}{c}\pm mc,
    \end{equation}
    they can be written as
    \begin{align}
    	(\alpha+k-n)C_k^1-\frac{i}{\hbar}
    	\left(p_zC_{k-1}^3+\varepsilon_+C_{k-1}^4\right)
    	&=0, \\
    	(\alpha+k+n+1)C_k^2+\frac{i}{\hbar}
    	\left(p_zC_{k-1}^4-\varepsilon_+C_{k-1}^3\right)
    	&=0, \\
    	(\alpha+k-n)C_k^3-\frac{i}{\hbar}
    	\left(p_zC_{k-1}^1+\varepsilon_-C_{k-1}^2\right)
    	&=0, \\
    	(\alpha+k+n+1)C_k^4+\frac{i}{\hbar}
    	\left(p_zC_{k-1}^2-\varepsilon_-C_{k-1}^1\right)
    	&=0.
    \end{align}

    These relations may be simplified into the proportionality structure
    \begin{align}
    	\frac{C_{k+1}^2}{C_k^1}
    	&=
    	-\frac{i(\lambda p_z-\varepsilon_+)}
    	{\hbar\lambda(\alpha+k+n+2)},\\
    	\frac{C_{k+1}^4}{C_k^1}
    	&=
    	-\frac{i(p_z-\lambda\varepsilon_-)}
    	{\hbar\lambda(\alpha+k+n+2)},\\
    	\frac{C_k^1}{C_{k-2}^1}
    	=
    	\frac{C_k^3}{C_{k-2}^3}
    	&=
    	-\frac{p_\kappa^2}
    	{\hbar^2(\alpha+k+n)(\alpha+k-n)},\\
    	\frac{C_{k+1}^2}{C_{k-1}^2}
    	=
    	\frac{C_{k+1}^4}{C_{k-1}^4}
    	&=
    	-\frac{p_\kappa^2}
    	{\hbar^2(\alpha+k+n+2)(\alpha+k-n)}.
    \end{align}
    The first two relations hold for $k\ge0$, the third for $k\ge2$, and the fourth for $k\ge1$.
    where
    \begin{equation}
    	p_\kappa^2=\frac{E^2}{c^2}-m^2c^2-p_z^2
    	\label{eq:pkappa}
    \end{equation}
    is the squared transverse momentum, and the ratio
    \begin{equation}
    	\lambda=\frac{C_k^1}{C_k^3}
    	\label{eq:lambda-def}
    \end{equation}
    is independent of $k$.
    
    The indicial equation has two roots. To ensure regularity of the spinor at the origin, one chooses
    \begin{equation}
    	\alpha=n\qquad (n\ge 0),
    	\qquad
    	\alpha=-n-1\qquad (n<0).
    	\label{eq:alpha-roots}
    \end{equation}
    These two roots have the same status as the two admissible radial indices in ordinary cylindrical-wave problems: they encode the local regularity structure of the series. They should not be confused with the genuinely new internal branch structure that will emerge only later.
    
    Since the two radial branches lead to the same subsequent spectral analysis, it is sufficient to focus on the regular case
    \begin{equation}
    	n\ge 0,\qquad \alpha=n.
    \end{equation}
    In this case, all odd terms vanish in the series for $R_1$ and $R_3$, whereas all even terms vanish in the series for $R_2$ and $R_4$. Setting $k=2m$, one obtains
    \begin{equation}
    	C_{2m}^{1}
    	=
    	\frac{\kappa^{2m}(-1)^m}
    	{2^{2m}m!\,(n+1)(n+2)\cdots(n+m)}\,C_0,
    \end{equation}
    where $C_0$ is a constant and
    \begin{equation}
    	\kappa=\frac{p_\kappa}{\hbar}.
    \end{equation}
    The radial function $R_1(r)$ therefore takes the form
    \begin{equation}
    	R_1(r)
    	=
    	C_0 r^n\sum_{m=0}^{\infty}
    	\frac{(-1)^m(\kappa r)^{2m}}
    	{2^{2m}m!\,(n+1)(n+2)\cdots(n+m)},
    	\label{eq:R1-series}
    \end{equation}
    where the product in the denominator is understood as unity for $m=0$. Comparing with the standard Bessel expansion
    \begin{equation}
    	J_n(\kappa r)
    	=
    	\sum_{m=0}^{\infty}
    	\frac{(-1)^m}{m!\Gamma(m+n+1)}
    	\left(\frac{\kappa r}{2}\right)^{2m+n},
    \end{equation}
    one identifies the exact radial dependence as Bessel functions. Once one spinor component is fixed, the remaining components follow from the recurrence relations. Define
    \begin{align}
    	g_\lambda&=-\frac{i}{\hbar\kappa}
    	\left(p_z-\frac{E+mc^2}{c\lambda}\right),
    	\nonumber\\
    	h_\lambda&=-\frac{i}{\hbar\kappa}
    	\left(\frac{p_z}{\lambda}-\frac{E-mc^2}{c}\right).
    	\label{eq:gh-SI}
    \end{align}
    The final exact cylindrical spinor can then be written compactly as
    \begin{equation}
    	|\psi_{n,\kappa,k_z}\rangle
    	=
    	e^{ip_z z/\hbar}e^{in\theta}
    	\begin{bmatrix}
    		J_n(\kappa r)\\
    		g_\lambda J_{n+1}(\kappa r)e^{i\theta}\\
    		\lambda^{-1}J_n(\kappa r)\\
    		h_\lambda J_{n+1}(\kappa r)e^{i\theta}
    	\end{bmatrix}.
    	\label{eq:wavefunction0}
    \end{equation}

    Equation~\eqref{eq:wavefunction0} is the exact unified free cylindrical Dirac family that will underlie the remainder of the paper. It is already an exact solution of Eq.~\eqref{eq:ham}; the relation of $\lambda$ to the standard cylindrical bases will be made explicit below.
    
    \subsection{The free parameter $\lambda$}
    \label{subsec:lambda-family}

    At fixed $(E,\kappa,k_z,J_z)$, the regular positive-energy cylindrical solution space is two-dimensional. The general solution in Eq.~\eqref{eq:wavefunction0} therefore contains a single free complex parameter $\lambda$. Its explicit relation to the conventional spin-polarized basis is derived in Sec.~\ref{sec:branch-structure}, after both basis modes have been introduced.

    The generalized series construction yields the full regular cylindrical family without choosing a preferred internal basis. Different exact bases correspond to particular values of $\lambda$ or to linear combinations of members of the same family. The transverse operator derived in Sec.~\ref{sec:K-operator} provides a conserved resolution of this two-dimensional freedom.

    \subsection{Previously known solutions as special cases of the unified family}
    \label{subsec:known-solutions}
    
    The generalized $\lambda$-family contains exact cylindrical modes used in several different relativistic settings. We first recover the conventional spin-polarized pair of Bliokh \textit{et al.}~\cite{Bliokh2011}. We then show that the cylindrical separation modes employed by Mihaila, Dawson, and Cooper in longitudinally boost-invariant Dirac-field quantization are special members of the same family~\cite{Mihaila2006}. Finally, helicity-adapted twisted states are related to this two-dimensional space by a unitary basis change~\cite{Serbo2015TwistedScattering}.
    
    \subsubsection{Reduction to the conventional spin-polarized vortex basis}
    \label{subsubsec:Bliokh-reduction}
    
    We begin from the general positive-energy cylindrical spinor in the Dirac representation. In natural units $\hbar=c=1$, it may be written as
    \begin{equation}
    	\psi_D
    	=
    	\mathcal N\,e^{i(k_z z+n\theta)}
    	\begin{pmatrix}
    		J_n(\kappa r)\\[1mm]
    		g\,J_{n+1}(\kappa r)e^{i\theta}\\[1mm]
    		\lambda^{-1}J_n(\kappa r)\\[1mm]
    		h\,J_{n+1}(\kappa r)e^{i\theta}
    	\end{pmatrix},
    	\label{eq:psiD-general}
    \end{equation}
    where
    \begin{align}
    	g&=-\frac{i}{\kappa}
    	\left(k_z-\frac{E+m}{\lambda}\right),
    	\nonumber\\
    	h&=-\frac{i}{\kappa}
    	\left(\frac{k_z}{\lambda}-(E-m)\right).
    	\label{eq:gh-def}
    \end{align}
    and
    \begin{equation}
    	E^2=k_z^2+\kappa^2+m^2.
    	\label{eq:dispersion-general}
    \end{equation}
    
    If we choose
    \begin{equation}
    	\lambda=\frac{E+m}{k_z},
    	\label{eq:lambda-Bliokh}
    \end{equation}
    which in ordinary units is
    \begin{equation}
    	\lambda=\frac{E+mc^2}{c p_z},
    \end{equation}
    then Eq.~\eqref{eq:gh-def} gives
    \begin{equation}
    	g=0,
    	\qquad
    	\lambda^{-1}=\frac{k_z}{E+m},
    	\qquad
    	h=\frac{i\kappa}{E+m}.
    	\label{eq:gh-Bliokh}
    \end{equation}
    Substituting Eq.~\eqref{eq:gh-Bliokh} into Eq.~\eqref{eq:psiD-general}, one obtains
    \begin{equation}
    	\psi_D
    	=
    	\mathcal N\,e^{i(k_z z+n\theta)}
    	\begin{pmatrix}
    		J_n(\kappa r)\\[1mm]
    		0\\[1mm]
    		\dfrac{k_z}{E+m}J_n(\kappa r)\\[3mm]
    		\dfrac{i\kappa}{E+m}J_{n+1}(\kappa r)e^{i\theta}
    	\end{pmatrix}.
    	\label{eq:psiD-Bliokh-up}
    \end{equation}
    Up to the overall normalization convention, this is precisely the spin-polarized cylindrical vortex mode used in the relativistic electron vortex-beam literature \cite{Bliokh2011}. Thus one member of the conventional spin-polarized REVB basis is obtained at the finite value in Eq.~\eqref{eq:lambda-Bliokh}. The second independent polarization mode is obtained from the limit $\lambda\to0$ after rescaling the overall normalization. Together, the two conventional modes span the same regular cylindrical solution space.
    
    Thus the generalized $\lambda$-family does not merely reproduce one convenient vortex mode: it displays the complete regular two-dimensional cylindrical solution space in which the conventional REVB pair is one basis. Section~\ref{sec:branch-structure} will show that the $K$ eigenstates provide a different basis of the same space.

    \subsubsection{Mihaila--Dawson--Cooper modes as special members of the general family}
    \label{subsubsec:MDC-reduction}

    The same family also contains the cylindrical modes used by Mihaila, Dawson, and Cooper (MDC) in the canonical quantization of a free Dirac field in longitudinally boost-invariant coordinates with azimuthal symmetry~\cite{Mihaila2006}. Since Ref.~\cite{Mihaila2006} employs a different metric, gamma-matrix representation, and Fourier convention, we state the comparison explicitly.

    The MDC paper uses
    \begin{equation}
        \eta_{\mu\nu}^{\rm MDC}=\operatorname{diag}(-1,1,1,1),
        \qquad
        (\gamma_{\rm MDC}^{\mu}\partial_{\mu}+m)\psi_{\rm MDC}=0,
        \label{eq:MDC-convention}
    \end{equation}
    with chiral gamma matrices, whereas Eq.~\eqref{eq:ham} is written in the standard Dirac-Pauli representation with signature $(+,-,-,-)$. The two conventions are related by
    \begin{equation}
        \begin{aligned}
        i\gamma_{\rm MDC}^{\mu}
        &=U\gamma_D^{\mu}U^{\dagger},
        &\qquad
        \psi_{\rm MDC}&=U\psi_D,\\
        U&=\frac{1}{\sqrt2}
        \begin{pmatrix}
            iI&-iI\\
            I&I
        \end{pmatrix}.
        \end{aligned}
        \label{eq:Dirac-to-MDC}
    \end{equation}
    Restoring the time dependence, our positive-energy mode is proportional to $e^{-iEt+ik_z z}$, while Ref.~\cite{Mihaila2006} uses $e^{iE_{\rm M}t-ik_{z,{\rm M}}z}$. The same mode is therefore compared with
    \begin{equation}
        E_{\rm M}=-E,
        \qquad
        k_{z,{\rm M}}=-k_z.
        \label{eq:MDC-Fourier-map}
    \end{equation}

    To avoid confusing their transverse separation eigenvalue with the parameter used in our general solution, define
    \begin{equation}
        \Lambda=\sqrt{m^2+\kappa^2},
        \qquad
        \ell_s=s\Lambda,
        \qquad s=\pm1.
        \label{eq:MDC-transverse-scale}
    \end{equation}
    Ref.~\cite{Mihaila2006} denotes the quantity $\ell_s$ by $\lambda$ and obtains the two values $\ell_s=\pm\Lambda$. In the fixed Dirac-frame convention $\lambda=C^1/C^3$ used in Eq.~\eqref{eq:lambda-def}, the corresponding two parameter values are
    \begin{equation}
        \lambda_{{\rm MDC},s}^{(D)}
        =-
        \frac{k_z-E+s\Lambda}{k_z-E-s\Lambda}.
        \label{eq:lambda-MDC-Dirac-chart}
    \end{equation}
    Because $\lambda$ is defined as a ratio of basis coefficients, a fixed rephasing of the second basis mode changes it according to $\widetilde\lambda=-\lambda^{(D)}$. In the MDC-adapted convention the same two modes are represented by
    \begin{equation}
        \widetilde\lambda_{{\rm MDC},s}
        =
        \frac{k_z-E+s\Lambda}{k_z-E-s\Lambda}
        .
        \label{eq:lambda-MDC-choice}
    \end{equation}
    In particular, the $s=+1$ value is the special parameter
    \begin{equation}
        \widetilde\lambda_{{\rm MDC},+}
        =
        \frac{k_z-E+\Lambda}{k_z-E-\Lambda}.
        \label{eq:lambda-MDC-plus}
    \end{equation}
    The relative sign between Eqs.~\eqref{eq:lambda-MDC-Dirac-chart} and \eqref{eq:lambda-MDC-choice} is therefore a convention in the definition of the coefficient ratio, not a difference between physical spinors.

    Substituting Eq.~\eqref{eq:lambda-MDC-Dirac-chart} into Eq.~\eqref{eq:psiD-general}, applying Eq.~\eqref{eq:Dirac-to-MDC}, and defining $j=n+\tfrac12$, introduce the compact radial-angular functions
    \begin{equation}
        f_- = e^{-i\theta/2}J_{j-1/2}(\kappa r),
        \qquad
        f_+ = e^{i\theta/2}J_{j+1/2}(\kappa r).
        \label{eq:MDC-fpm}
    \end{equation}
    Up to an overall normalization and phase, the transformed spinor is
    \begin{equation}
        \psi_{\rm MDC}^{(s)}
        =\mathcal N_s e^{i(E_{\rm M}t+j\theta-k_{z,{\rm M}}z)}
        \begin{pmatrix}
        -f_-\\[1mm]
        \dfrac{\kappa}{\ell_s-m}
        \dfrac{i\ell_s}{k_{z,{\rm M}}-E_{\rm M}}f_+\\[3mm]
        -\dfrac{i\ell_s}{k_{z,{\rm M}}-E_{\rm M}}f_-\\[3mm]
        \dfrac{\kappa}{\ell_s-m}f_+
        \end{pmatrix}.
        \label{eq:psi-MDC-final}
    \end{equation}
    This is the cylindrical free-field spinor of Ref.~\cite{Mihaila2006} (their Eq.~(4.36)), after the convention map in Eqs.~\eqref{eq:MDC-convention}--\eqref{eq:MDC-Fourier-map}. Thus the MDC modes are not a separate solution family: both transverse separation branches are contained in the general regular $\lambda$-family. The operator interpretation of the MDC separation label is not needed for this embedding and is left outside the scope of the present paper.

    \subsubsection{Helicity-adapted twisted states as a basis realization}
    \label{subsubsec:helicity-remark}
    
    The helicity-adapted twisted states used in non-plane-wave scattering theory form another basis of the same positive-energy cylindrical Dirac subspace rather than an independent solution family. In the conventional vortex construction of Ref.~\cite{Bliokh2011}, the polarization spinor may be chosen in either the $\sigma_z$ basis or the helicity basis. Since the spin-polarized vortex basis is contained in the present $\lambda$-family, the helicity-adapted states are obtained by a basis transformation within the same free cylindrical sector.
    
    The physical distinction is therefore not at the level of the solution space itself, but in the operator chosen to resolve its twofold freedom. The conventional construction uses a fixed canonical-polarization basis, the helicity construction diagonalizes helicity, and the $K$ basis introduced below diagonalizes a purely transverse conserved operator. These are different exact resolutions of the same cylindrical Dirac sector rather than independent solution families.
    
    The spin-polarized vortex pair, the MDC separation modes, and the helicity-adapted twisted states therefore belong to one common regular cylindrical solution space, although they use different representations or internal bases. The parameter $\lambda$ fixes the relative coefficients within this family. In the next section we introduce a conserved transverse operator whose eigenstates correspond to two distinguished members of the family.
	
	\section{Transverse conserved operator and symmetry-adapted branches}
	\label{sec:K-operator}
	
	The generalized $\lambda$-family retains a two-dimensional freedom at fixed cylindrical quantum numbers. An additional conserved operator provides a purely transverse resolution of this degeneracy.
	
	The operator is derived using the covariant tetrad formulation of the Dirac equation in flat spacetime expressed in cylindrical coordinates \cite{Weyl1929}. The role of the vierbein is purely kinematical: it provides a convenient local frame in which the curvilinear-coordinate Dirac equation can be reorganized into a form suitable for separation. Throughout this section we work in natural units, $\hbar=c=1$.
	
	\subsection{Derivation of the transverse conserved operator}
	\label{subsec:K-derivation}
	
	Let $x^\mu=(t,r,\theta,z)$ and
	\begin{equation}
		ds^2 = dt^2-dr^2-r^2 d\theta^2-dz^2 .
		\label{eq:cyl-metric}
	\end{equation}
	A natural orthonormal tetrad is
	\begin{equation}
		e^a{}_{\mu}=\mathrm{diag}(1,1,r,1),
		\qquad
		e_a{}^{\mu}=\mathrm{diag}\!\left(1,1,\frac1r,1\right).
		\label{eq:tetrad}
	\end{equation}
	The covariant Dirac equation in flat spacetime, expressed in this curvilinear frame, reads
	\begin{equation}
		\left[
		i\gamma^a e_a{}^{\mu}\left(\partial_\mu+\Gamma_\mu\right)-m
		\right]\psi=0,
		\label{eq:covariant-dirac}
	\end{equation}
	where $\Gamma_\mu$ is the spin connection associated with the cylindrical tetrad.
	
	It is convenient to introduce the $\theta$-dependent similarity transformation
	\begin{equation}
		S(\theta)=\exp\!\left(-\frac{\theta}{2}\gamma^1\gamma^2\right),
		\qquad
		\psi_d=S^{-1}\psi ,
		\label{eq:similarity-S}
	\end{equation}
	which rotates the local frame into the diagonal cylindrical gauge. After standard algebra \cite{Villalba1989}, the Dirac equation takes the form
	\begin{equation}
		\left[
		i\gamma^0\partial_t
		+i\gamma^1\left(\partial_r+\frac{1}{2r}\right)
		+i\gamma^2\frac{1}{r}\partial_\theta
		+i\gamma^3\partial_z
		-m
		\right]\psi_d=0 .
		\label{eq:dirac-local-frame}
	\end{equation}
	Equation~\eqref{eq:dirac-local-frame} is still the free Dirac equation in flat spacetime; the tetrad formalism has merely reorganized the cylindrical-coordinate kinematics into a form adapted to separation.
	
	To isolate the residual cylindrical structure, we set
	\begin{equation}
		\psi_d=\gamma^3\gamma^0\,\Phi .
		\label{eq:psi-gammaPhi}
	\end{equation}
	Equation~\eqref{eq:dirac-local-frame} can then be written as
	\begin{equation}
		\left(\hat K_1+\hat K_2\right)\Phi=0,
		\label{eq:K1K2sum}
	\end{equation}
	with
	\begin{align}
		\hat K_1
		&=
		\left[
		\gamma^1\left(\partial_r+\frac{1}{2r}\right)
		+\gamma^2\frac{1}{r}\partial_\theta
		\right]\gamma^3\gamma^0,
		\label{eq:K1-def}
		\\
		\hat K_2
		&=
		\left[
		\gamma^0\partial_t+\gamma^3\partial_z+i m
		\right]\gamma^3\gamma^0.
		\label{eq:K2-def}
	\end{align}
	Because the derivatives in the two sets of variables commute and the accompanying Dirac matrices satisfy the Clifford algebra, a direct calculation gives $[\hat K_1,\hat K_2]=0$. One may therefore introduce a separation constant $\lambda'$ through
	\begin{align}
		\hat K_1\Phi &= \lambda' \Phi,
		\label{eq:K1-eig}
		\\
		\hat K_2\Phi &= -\lambda' \Phi.
		\label{eq:K2-eig}
	\end{align}
	
	For stationary states with definite longitudinal momentum,
	\begin{equation}
		\Phi(t,r,\theta,z)=e^{-iEt+i k_z z}\,\phi(r,\theta),
		\label{eq:stationary-ansatz-K}
	\end{equation}
	Eq.~\eqref{eq:K2-eig} becomes purely algebraic. Using the Clifford algebra, one finds
	\begin{equation}
		\hat K_2^{\,2}\Phi
		=
		\left(E^2-k_z^2-m^2\right)\Phi.
		\label{eq:K2-square}
	\end{equation}
	Defining the transverse momentum by
	\begin{equation}
		p_\kappa^2 \equiv E^2-m^2-k_z^2,
		\label{eq:pkappa-def-K}
	\end{equation}
	one obtains
	\begin{equation}
		\lambda'=\pm p_\kappa .
		\label{eq:lambdaprime-pkappa}
	\end{equation}
	
	The separated transverse operator in the original cylindrical spinor basis is obtained by transforming back,
	\begin{equation}
		\widetilde K = S\,\hat K_1\,S^{-1}.
		\label{eq:K-from-K1}
	\end{equation}
	The transformation gives
	\begin{equation}
	\begin{aligned}
		\widetilde K={}&
		\gamma^1\gamma^0\gamma^3
		\left(
		\cos\theta\,\partial_r
		-\frac{\sin\theta}{r}\partial_\theta
		\right)
		\\
		&+
		\gamma^2\gamma^0\gamma^3
		\left(
		\sin\theta\,\partial_r
		+\frac{\cos\theta}{r}\partial_\theta
		\right).
	\end{aligned}
		\label{eq:K-differential}
	\end{equation}
	Using $\mathbf p=-i\boldsymbol{\nabla}$, the differential operator in Eq.~\eqref{eq:K-differential} is
	\begin{equation}
		\widetilde K=-\beta(\boldsymbol{\Sigma}\times \mathbf p)_z.
	\end{equation}
	The overall sign of a separation operator is conventional and only exchanges its two eigenvalue labels. For later convenience we therefore define
	\begin{equation}
		K\equiv-\widetilde K
		=\beta(\boldsymbol{\Sigma}\times \mathbf p)_z,
		\label{eq:K-cartesian-compact}
	\end{equation}
	and use this sign convention throughout the remainder of the paper.

	The operator in Eq.~\eqref{eq:K-cartesian-compact} is Hermitian on the usual momentum-operator domain, is purely transverse, and commutes with the operators that label the free cylindrical problem,
	\begin{equation}
		[K,H]=0,
		\qquad
		[K,p_z]=0,
		\qquad
		[K,J_z]=0,
		\label{eq:K-commutators}
	\end{equation}
	where
	\begin{equation}
		H=\boldsymbol{\alpha}\cdot\mathbf p+\beta m,
		\qquad
		J_z=-i\partial_\theta+\frac12\Sigma_z.
		\label{eq:H-pz-Jz-def}
	\end{equation}
	Moreover,
	\begin{equation}
		K^2=p_x^2+p_y^2\equiv p_\perp^2,
		\label{eq:K-square-operator}
	\end{equation}
	so that on states of definite transverse momentum,
	\begin{equation}
		K^2\psi=p_\kappa^2\psi.
		\label{eq:K-square-eigen}
	\end{equation}
	For $p_\perp>0$, the normalized operator $\Gamma_K=K/\sqrt{p_\perp^2}$ has eigenvalues $\eta=\pm1$. At $p_\perp=0$ the two eigenvalues merge, and this branch normalization is not defined.
	
	\subsection{Fixing $\lambda$ and constructing the two branch eigenstates}
	\label{subsec:K-branches}
	
	Consider the general cylindrical family derived in Sec.~\ref{sec:generalized-series},
	\begin{equation}
		\psi_\lambda
		=
		N e^{ik_z z}e^{in\theta}
		\begin{bmatrix}
			J_n(\kappa r)\\[1mm]
			-\dfrac{i}{\kappa}
			\left(
			k_z-\dfrac{E+m}{\lambda}
			\right)
			J_{n+1}(\kappa r)e^{i\theta}\\[3mm]
			\dfrac{1}{\lambda}J_n(\kappa r)\\[3mm]
			-\dfrac{i}{\kappa}
			\left(
			\dfrac{k_z}{\lambda}-(E-m)
			\right)
			J_{n+1}(\kappa r)e^{i\theta}
		\end{bmatrix},
		\label{eq:psi-lambda-general-Ksec}
	\end{equation}
	where
	\begin{equation}
		E^2=m^2+k_z^2+\kappa^2.
		\label{eq:dispersion-Ksec}
	\end{equation}
	Substituting Eq.~\eqref{eq:psi-lambda-general-Ksec} into the eigenvalue equation
	\begin{equation}
		K\psi=\lambda'\psi
		\label{eq:K-eigenvalue-eq}
	\end{equation}
	fixes the previously free parameter $\lambda$. The two allowed values are
	\begin{equation}
		\lambda_\pm=\frac{k_z\pm i\kappa}{E-m},
		\label{eq:lambda-pm-final}
	\end{equation}
	corresponding to
	\begin{equation}
		\lambda'=\pm p_\kappa = \pm \kappa
		\qquad
		(\hbar=c=1).
		\label{eq:lambdaprime-pm-final}
	\end{equation}
	
	We denote the two symmetry-adapted branch eigenstates by $\psi_{K+}$ and $\psi_{K-}$, corresponding to the eigenvalues $K=+p_\kappa$ and $K=-p_\kappa$, respectively. The two positive-energy $K$-eigenstates are therefore
	\begin{equation}
		\psi_{K+}
		=
		N e^{ik_z z}e^{in\theta}
		\begin{bmatrix}
			J_n(\kappa r)\\[1mm]
			J_{n+1}(\kappa r)e^{i\theta}\\[2mm]
			\dfrac{k_z-i\kappa}{E+m}J_n(\kappa r)\\[3mm]
			-\dfrac{k_z-i\kappa}{E+m}J_{n+1}(\kappa r)e^{i\theta}
		\end{bmatrix},
		\label{eq:psiKplus-final}
	\end{equation}
	and
	\begin{equation}
		\psi_{K-}
		=
		N e^{ik_z z}e^{in\theta}
		\begin{bmatrix}
			J_n(\kappa r)\\[1mm]
			-\,J_{n+1}(\kappa r)e^{i\theta}\\[2mm]
			\dfrac{k_z+i\kappa}{E+m}J_n(\kappa r)\\[3mm]
			\dfrac{k_z+i\kappa}{E+m}J_{n+1}(\kappa r)e^{i\theta}
		\end{bmatrix}.
		\label{eq:psiKminus-final}
	\end{equation}
	These are the two exact cylindrical branches selected by the transverse conserved operator $K$ in the sign convention of Eq.~\eqref{eq:K-cartesian-compact}. Component by component, Eqs.~\eqref{eq:psiKplus-final} and \eqref{eq:psiKminus-final} reproduce, up to normalization and convention changes, the positive-energy transverse-helicity cylindrical spinors used in earlier quantization, rotating-frame, and confined-Dirac analyses~\cite{BalantekinDeWeerd1995,JiangLiao2016,Khosravi2019PRB}. The present derivation obtains this branch resolution from the general regular free family and makes its relation to the other cylindrical bases explicit.
	
	The normalization constant depends on the chosen longitudinal and radial regularization. For box normalization along $z$ with length $D$ and finite-radius regularization $R$, one may write
	\begin{align}
			N&=
			\sqrt{
				\frac{E+m}{4\pi E\,D\,I_1(R)}
			},
			\nonumber\\
			I_1(R)&=\int_0^R
			\left[
			J_n^2(\kappa r)+J_{n+1}^2(\kappa r)
			\right]r\,dr.
			\label{eq:N-regularized}
		\end{align}
	The ideal cylindrical limit is recovered by taking $R\to\infty$ only after forming properly normalized expectation values. In practical numerical calculations, a finite effective radius may be introduced as a regularization of the ideal Bessel state.
	
	Equations~\eqref{eq:psiKplus-final} and \eqref{eq:psiKminus-final} give a transverse-symmetry resolution of the generalized family. The operator $K$ supplies an additional good quantum number for the free cylindrical problem, while other conserved bases, such as helicity, provide different resolutions of the same two-dimensional space.
	
\section{Relation to standard cylindrical bases}
\label{sec:branch-structure}

At fixed $(E,\kappa,k_z,J_z)$ the positive-energy cylindrical sector is two-dimensional. The $K$ basis is therefore related by a unitary transformation to any other complete basis of that sector. This equivalence does not remove the usefulness of $K$: it identifies which linear combinations diagonalize a purely transverse conserved operator.

\subsection{Conventional spin-polarized modes}

A standard pair of cylindrical vortex modes is
\begin{align}
\psi_{\uparrow}
&=N e^{ik_z z}e^{in\theta}
\begin{bmatrix}
J_n\\ 0\\
\dfrac{k_z}{E+m}J_n\\
\dfrac{i\kappa}{E+m}J_{n+1}e^{i\theta}
\end{bmatrix},
\label{eq:psi-up-spinbasis}
\\[1ex]
\psi_{\downarrow}
&=N e^{ik_z z}e^{in\theta}
\begin{bmatrix}
0\\ J_{n+1}e^{i\theta}\\
-\dfrac{i\kappa}{E+m}J_n\\
-\dfrac{k_z}{E+m}J_{n+1}e^{i\theta}
\end{bmatrix},
\label{eq:psi-down-spinbasis}
\end{align}
where the Bessel-function arguments are $\kappa r$. Direct comparison with the general mode in Eq.~\eqref{eq:psiD-general} gives, for finite $\lambda$ and the same overall normalization,
\begin{equation}
\psi_{\lambda}
=
\psi_{\uparrow}+\zeta(\lambda)\psi_{\downarrow},
\qquad
\zeta(\lambda)=g
=-\frac{i}{\kappa}\left(k_z-\frac{E+m}{\lambda}\right).
\label{eq:zeta-lambda}
\end{equation}
The second component fixes $\zeta=g$. The remaining two nontrivial components then agree because
\begin{align}
k_z-i\kappa g
&=\frac{E+m}{\lambda},\nonumber\\
i\kappa-k_zg
&=(E+m)h,
\label{eq:zeta-component-check}
\end{align}
where the second identity uses $E^2=m^2+\kappa^2+k_z^2$. Thus $\lambda$ and $\zeta$ are simply two equivalent ways of parameterizing the same general linear combination. Inverting Eq.~\eqref{eq:zeta-lambda} gives
\begin{equation}
\lambda=\frac{E+m}{k_z-i\kappa\zeta}.
\label{eq:lambda-zeta-inverse}
\end{equation}

With the phase convention used above,
\begin{equation}
\psi_{K+}=\frac{\psi_{\uparrow}+\psi_{\downarrow}}{\sqrt2},
\qquad
\psi_{K-}=\frac{\psi_{\uparrow}-\psi_{\downarrow}}{\sqrt2}.
\label{eq:K-Hadamard-basis}
\end{equation}
Thus the conventional modes are equal-weight superpositions of the two $K$ branches, and conversely.

The labels $\uparrow$ and $\downarrow$ refer to the constant rest-frame polarization spinors used to construct the Bessel modes~\cite{Bliokh2011}. They should not be confused with eigenvalues of the bare four-component operator $\Sigma_z$ for the full nonparaxial mode.

\subsection{Helicity and angular momentum}

Helicity,
\begin{equation}
 h=\frac{\boldsymbol{\Sigma}\cdot\mathbf p}{|\mathbf p|},
 \qquad |\mathbf p|=\sqrt{\kappa^2+k_z^2},
\end{equation}
is also conserved for the free Dirac Hamiltonian. Helicity-adapted twisted states are obtained by the usual momentum-dependent rotation of the rest-frame polarization spinor~\cite{IvanovSerbo2011,Serbo2015TwistedScattering}. They therefore constitute another exact basis of the same cylindrical sector, rather than a different solution family.

The distinction between the $K$ and helicity resolutions is captured by the operator algebra
\begin{equation}
 \{K,h\}=0,
 \qquad
 \{K,\Sigma_z\}=0.
 \label{eq:K-anticommutators}
\end{equation}
For a normalized $K$ eigenstate with $\kappa>0$, Eq.~\eqref{eq:K-anticommutators} implies
\begin{equation}
 \langle h\rangle_{K\eta}=0,
 \qquad
 \langle \Sigma_z\rangle_{K\eta}=0.
 \label{eq:zero-helicity-spin}
\end{equation}
Equivalently, a $K$ branch contains equal probabilities of the two helicities, although their relative phase depends on momentum and convention. Since every component of Eqs.~\eqref{eq:psiKplus-final} and \eqref{eq:psiKminus-final} has the same total angular momentum,
\begin{equation}
 J_z\psi_{K\eta}=\left(n+\frac12\right)\psi_{K\eta},
\end{equation}
Eq.~\eqref{eq:zero-helicity-spin} yields, in the ideal cylindrical normalization,
\begin{equation}
 \langle S_z\rangle_{K\eta}=0,
 \qquad
 \langle L_z\rangle_{K\eta}=\left(n+\frac12\right)\hbar.
\end{equation}
This result is consistent with the conventional spin-orbit and angular-momentum descriptions of relativistic vortex modes~\cite{Bliokh2011,Bliokh17pra,Bliokh2017Review}, while showing that the same cylindrical sector admits a transverse branch basis with vanishing longitudinal spin expectation value.

\section{Symmetry-resolved quantum geometry of the $K$ branches}
\label{sec:quantum-geometry}

The geometric analysis in this section follows from the symmetry resolution obtained above. At a single momentum point, the positive-energy sector is a two-dimensional degenerate eigenspace, and no unique Abelian geometry can be attached to an arbitrary choice of basis within it. The conserved operator $K$ resolves this sector into two symmetry-defined one-dimensional eigenspaces. The geometric question then concerns the variation of these eigenspaces as $(\kappa,k_z)$ changes. The free dispersion alone does not characterize this internal variation, because it is identical for the two branches. Instead, each branch is represented by a rank-one joint spectral projector whose parameter derivatives measure how the corresponding subspace changes in Hilbert space. The quantum geometric tensor is the gauge-invariant differential object that encodes this change, with its symmetric and antisymmetric parts giving the quantum metric and Berry curvature, respectively. Thus the geometry considered below is induced by the symmetry-selected branch decomposition rather than by an arbitrary basis choice in the unresolved doublet. We work at a representative momentum azimuth,
\begin{equation}
 \mathbf p=(\kappa,0,k_z),
 \qquad
 E=\sqrt{m^2+\kappa^2+k_z^2},
 \label{eq:representative-momentum}
\end{equation}
which is sufficient because changing the azimuth acts by a common axial rotation and leaves the scalar geometric invariants below unchanged.

The positive-energy projector is
\begin{equation}
 P_+(\kappa,k_z)
 =\frac12\left(1+\frac{H(\kappa,k_z)}{E}\right),
 \label{eq:positive-energy-projector}
\end{equation}
and on sectors with $\kappa>0$ we define
\begin{equation}
 \Gamma_K=\frac{K}{\kappa},
 \qquad
 \Gamma_K^2=1,
 \qquad
 [P_+,\Gamma_K]=0.
 \label{eq:normalized-K}
\end{equation}
The two rank-one branch projectors are then
\begin{equation}
 \Pi_\eta
 =P_+\frac{1+\eta\Gamma_K}{2},
 \qquad \eta=\pm1 .
 \label{eq:branch-projectors}
\end{equation}
Here $P_+$ first selects the two-dimensional positive-energy sector, while
\begin{equation}
 Q_\eta=\frac{1+\eta\Gamma_K}{2}
 \label{eq:K-spectral-projector}
\end{equation}
selects the eigenspace of $K$ with eigenvalue $\eta\kappa$, including both energy signs. Since $[P_+,Q_\eta]=0$, their product $\Pi_\eta=P_+Q_\eta$ is the joint rank-one projector onto the positive-energy $K_\eta$ branch. In particular,
\begin{equation}
 \Pi_++\Pi_-=P_+,
 \qquad
 \Pi_+\Pi_-=0.
 \label{eq:branch-projector-resolution}
\end{equation}
The joint projector is invariant under a local $U(1)$ rephasing of the corresponding branch eigenvector. The internal quantum metric and Berry curvature can therefore be written directly in terms of the projector~\cite{GrafPiechon2021},
\begin{align}
 g_{ij}^{(\eta)}
 &=\frac12\operatorname{Tr}\!\left[
 (\partial_i\Pi_\eta)(\partial_j\Pi_\eta)
 \right],
 \label{eq:projector-metric}
 \\
 \Omega_{ij}^{(\eta)}
 &=i\operatorname{Tr}\!\left[
 \Pi_\eta[\partial_i\Pi_\eta,\partial_j\Pi_\eta]
 \right],
 \label{eq:projector-curvature}
\end{align}
where $i,j\in\{\kappa,k_z\}$ and the trace is over the four Dirac components.

Evaluating Eqs.~\eqref{eq:projector-metric} and \eqref{eq:projector-curvature} gives
\begin{equation}
 \Omega_{\kappa k_z}^{(\eta)}
 =-\eta\frac{m}{2E^3}
 \label{eq:Berry-curvature-final}
\end{equation}
and
\begin{align}
 g_{\kappa\kappa}^{(\eta)}
 &=\frac{m^2+k_z^2}{4E^4},
 &
 g_{k_zk_z}^{(\eta)}
 &=\frac{m^2+\kappa^2}{4E^4},
 \nonumber\\
 g_{\kappa k_z}^{(\eta)}
 &=-\frac{\kappa k_z}{4E^4}.
 \label{eq:metric-final}
\end{align}
The two branches therefore have opposite Berry curvature and the same quantum metric. An equivalent calculation using the explicit two-component branch spinors gives the same result. In the projector formulation, invariance under a momentum-dependent rephasing of the branch eigenstate follows directly from the invariance of $\Pi_\eta$.

The metric and curvature are not independent. Their determinant satisfies
\begin{equation}
 \det g^{(\eta)}
 =g_{\kappa\kappa}^{(\eta)}g_{k_zk_z}^{(\eta)}
 -\left(g_{\kappa k_z}^{(\eta)}\right)^2
 =\frac{m^2}{16E^6},
 \label{eq:metric-determinant}
\end{equation}
while Eq.~\eqref{eq:Berry-curvature-final} gives
\begin{equation}
 \det g^{(\eta)}
 =\frac14\left(\Omega_{\kappa k_z}^{(\eta)}\right)^2 .
 \label{eq:metric-curvature-saturation}
\end{equation}
Hence the two-level metric--curvature inequality is saturated. In the massless limit the branch curvature vanishes and the metric becomes rank one, $\det g\to0$, even though individual metric components can remain finite.

For a closed loop $C$ lying entirely in the physical half-plane
\begin{equation}
 \mathcal B_{\rm cyl}=(0,\infty)\times\mathbb R,
\end{equation}
the geometric phase of a branch-preserving cycle is
\begin{equation}
 \gamma_\eta[C]
 =\int_{\Sigma_C}\Omega_{\kappa k_z}^{(\eta)}
 \,d\kappa\,dk_z,
 \label{eq:branch-Berry-phase}
\end{equation}
where $\partial\Sigma_C=C$. The relative phase between the two branches is therefore
\begin{equation}
 \Delta\gamma_K[C]
 =\gamma_+[C]-\gamma_-[C]
 =-\int_{\Sigma_C}\frac{m}{E^3}\,d\kappa\,dk_z .
 \label{eq:relative-branch-phase}
\end{equation}
For a sufficiently small loop of oriented area $\mathcal A_C$ centered at $(\kappa_0,k_{z0})$,
\begin{equation}
 \Delta\gamma_K[C]
 \simeq-\frac{m}{E_0^3}\mathcal A_C,
 \qquad
 E_0=\sqrt{m^2+\kappa_0^2+k_{z0}^2}.
 \label{eq:small-loop-phase}
\end{equation}
Because the two free branches are degenerate, Eqs.~\eqref{eq:branch-Berry-phase} and \eqref{eq:relative-branch-phase} apply to controlled evolutions that preserve the instantaneous $K$ decomposition. Degeneracy by itself does not suppress interbranch mixing, and the quantum metric does not determine a universal transition probability.

The geometry above is thus the symmetry-resolved internal geometry of the positive-energy cylindrical sector at fixed momentum azimuth. Its Abelian character is a consequence of resolving the rank-two positive-energy subspace with the conserved operator $K$; it is not associated with an arbitrary choice of basis inside the degenerate doublet. It does not include the additional parameter dependence of a finite-width or spatially regularized Bessel envelope. The parent rank-two positive-energy geometry remains non-Abelian and is not developed here.

\section{Symmetry protection and branch dynamics}
\label{sec:selection}

The operator $K$ labels the free eigenstates but does not enter the Hamiltonian as an energy-splitting term. Its dynamical implications follow from symmetry-preserving evolution, symmetry-breaking matrix elements, and transformations between the $K$ basis and conventional cylindrical modes. In this section we use natural units, $\hbar=c=1$.

\subsection{Symmetry protection and branch conversion}
\label{subsec:protection-conversion}

Let
\begin{equation}
 \Gamma_K=K\,(p_\perp^2)^{-1/2}
 \label{eq:GammaK-dynamics}
\end{equation}
be defined on sectors with $p_\perp>0$. If a time-dependent Hamiltonian obeys
\begin{equation}
 [H(t),\Gamma_K]=0
 \qquad \text{for all }t,
 \label{eq:instantaneous-Gamma-symmetry}
\end{equation}
then its evolution operator
\begin{equation}
 U(t,t_0)=\mathcal T\exp\!\left[-i\int_{t_0}^{t}H(t')\,dt'\right]
 \label{eq:time-evolution-operator}
\end{equation}
commutes with $\Gamma_K$. Differentiation gives
\begin{equation}
 \frac{d}{dt}\left(U^\dagger\Gamma_KU\right)
 =iU^\dagger[H(t),\Gamma_K]U=0,
 \label{eq:Gamma-conservation-proof}
\end{equation}
so that $U^\dagger\Gamma_KU=\Gamma_K$. The result is nonperturbative and remains valid for an arbitrarily strong time-dependent interaction satisfying Eq.~\eqref{eq:instantaneous-Gamma-symmetry}. The stronger condition $[H(t),K]=0$ conserves the full eigenvalue $\eta\kappa$, rather than only the branch sign $\eta$.

For branch eigenstates, define
\begin{equation}
 \mathcal M_{\eta_f\eta_i}
 =\langle f,\eta_f|U(t,t_0)|i,\eta_i\rangle .
 \label{eq:branch-amplitude}
\end{equation}
Equation~\eqref{eq:Gamma-conservation-proof} gives
\begin{equation}
 \mathcal M_{\eta_f\eta_i}=0
 \qquad (\eta_f\neq\eta_i)
 \label{eq:branch-selection}
\end{equation}
whenever the evolution preserves $\Gamma_K$. In terms of branch-resolved probabilities or cross sections,
\begin{align}
 \sigma_{\rm cons}
 &=\sigma_{+\to+}+\sigma_{-\to-},
 \nonumber\\
 \sigma_{\rm flip}
 &=\sigma_{+\to-}+\sigma_{-\to+},
 \label{eq:branch-cross-sections}
\end{align}
this yields the selection rule
\begin{equation}
 [U,\Gamma_K]=0
 \quad\Longrightarrow\quad
 \sigma_{\rm flip}=0.
 \label{eq:null-branch-flip}
\end{equation}
A complementary branch asymmetry,
\begin{equation}
 \mathcal A_K
 =\frac{\sigma_{+\to+}-\sigma_{-\to-}}
 {\sigma_{+\to+}+\sigma_{-\to-}},
 \label{eq:branch-asymmetry}
\end{equation}
compares the two branch-preserving responses. The branch-changing contribution $\sigma_{\rm flip}$ and the diagonal asymmetry $\mathcal A_K$ probe distinct components of the symmetry-resolved dynamics.

The distinction between symmetry-preserving and symmetry-breaking interactions can be made algebraically. For any perturbation $V$, define its $K$-even and $K$-odd parts by
\begin{equation}
 V_{\parallel}
 =\frac12\left(V+\Gamma_KV\Gamma_K\right),
 \qquad
 V_{\perp}
 =\frac12\left(V-\Gamma_KV\Gamma_K\right).
 \label{eq:V-parallel-perp}
\end{equation}
They satisfy
\begin{equation}
 [V_{\parallel},\Gamma_K]=0,
 \qquad
 \{V_{\perp},\Gamma_K\}=0.
 \label{eq:V-parity-relations}
\end{equation}
Writing $Q_\eta=(1+\eta\Gamma_K)/2$, or equivalently using the positive-energy branch projectors $\Pi_\eta=P_+Q_\eta$, one finds
\begin{equation}
 \Pi_{-\eta}V_{\parallel}\Pi_\eta=0,
 \qquad
 \Pi_\eta V_{\perp}\Pi_\eta=0.
 \label{eq:V-projector-selection}
\end{equation}
Accordingly, $V_{\parallel}$ acts within each branch, whereas $V_{\perp}$ contains the branch-off-diagonal matrix elements. This decomposition does not rely on a weak-coupling approximation.

For a scalar potential,
\begin{equation}
 H(t)=H_0+V(\mathbf r,t)I_4,
 \label{eq:scalar-potential-H}
\end{equation}
using $[p_i,V]=-i\partial_iV$ and
$K=\beta(\Sigma_xp_y-\Sigma_yp_x)$ gives
\begin{equation}
 [K,V]
 =-i\beta\left(
 \Sigma_x\partial_yV-\Sigma_y\partial_xV
 \right).
 \label{eq:K-V-general}
\end{equation}
A potential that is uniform in the transverse plane,
\begin{equation}
 V=V(z,t),
\end{equation}
therefore obeys
\begin{equation}
 [K,V(z,t)]=0.
 \label{eq:longitudinal-preserves-K}
\end{equation}
A longitudinal potential region with negligible transverse variation therefore preserves the full $K$ eigenvalue at arbitrary interaction strength and cannot induce branch conversion.

For an axisymmetric potential $V=V(r,z,t)$, introduce
\begin{equation}
 \Sigma_\theta=-\Sigma_x\sin\theta+\Sigma_y\cos\theta .
\end{equation}
Equation~\eqref{eq:K-V-general} becomes
\begin{equation}
 [K,V(r,z,t)]
 =i\beta\Sigma_\theta\,\partial_rV .
 \label{eq:K-V-radial}
\end{equation}
The transverse radial gradient therefore enters directly in the $K$-symmetry-breaking commutator. For free initial and final branch eigenstates,
\begin{equation}
 K|i,\eta_i\rangle=\eta_i\kappa_i|i,\eta_i\rangle,
 \qquad
 K|f,\eta_f\rangle=\eta_f\kappa_f|f,\eta_f\rangle,
\end{equation}
the exact identity
\begin{equation}
 \left(\eta_f\kappa_f-\eta_i\kappa_i\right)
 \langle f,\eta_f|V|i,\eta_i\rangle
 =\langle f,\eta_f|[K,V]|i,\eta_i\rangle
 \label{eq:commutator-matrix-element}
\end{equation}
relates every branch-changing matrix element to the symmetry-breaking commutator. In a fixed-$\kappa$ flip channel, $\eta_f=-\eta_i=-\eta$, this gives
\begin{equation}
 \langle f,-\eta|V|i,\eta\rangle
 =-\frac{i}{2\eta\kappa}
 \left\langle f,-\eta\left|
 \beta\Sigma_\theta\,\partial_rV
 \right|i,\eta\right\rangle .
 \label{eq:gradient-flip-matrix-element}
\end{equation}
The corresponding first-order interaction-picture amplitude is
\begin{equation}
 \mathcal A_{-\eta,\eta}^{(1)}(t)
 =-\frac{1}{2\eta\kappa}
 \int_{t_0}^{t}dt'\,
 \left\langle f,-\eta\left|
 \beta\Sigma_\theta\,\partial_rV_I(t')
 \right|i,\eta\right\rangle .
 \label{eq:first-order-gradient-amplitude}
\end{equation}
Because the two free $K$ branches are degenerate, no free-branch energy denominator appears. The conversion is determined by the projected $K$-odd matrix element and the interaction duration.

As an illustrative weakly inhomogeneous profile, consider
\begin{equation}
 V(r,z,t)=V_0(z,t)
 +\epsilon f(z,t)\frac{r^2}{2R^2}.
 \label{eq:weak-radial-profile}
\end{equation}
Then
\begin{equation}
 [K,V]
 =i\epsilon f(z,t)\frac{r}{R^2}\beta\Sigma_\theta,
 \label{eq:weak-radial-commutator}
\end{equation}
so that the branch-flip amplitude is linear in the transverse inhomogeneity $\epsilon$, whereas the conversion probability begins at order $\epsilon^2$. The proportionality coefficient depends on the radial mode profile and the interaction region; the scaling and the vanishing at $\epsilon=0$ are fixed by the symmetry.

When the interaction is sufficiently mode selective that transitions to other $(\kappa,k_z,J_z)$ sectors and to the negative-energy subspace are negligible, the dynamics closes within the selected doublet. Projecting onto the positive-energy doublet
\begin{equation}
 \mathcal H_K=\operatorname{span}\{|K+\rangle,|K-\rangle\},
\end{equation}
one may write the most general Hermitian effective Hamiltonian as
\begin{equation}
 H_{\rm eff}
 =(E+v_0)I_2
 +v_x\tau_x+v_y\tau_y+v_z\tau_z,
 \label{eq:effective-K-Hamiltonian}
\end{equation}
where $\tau_z$ has eigenstates $|K\pm\rangle$. In terms of the projected perturbation matrix elements,
\begin{align}
 2v_0
 &=\langle K+|V|K+\rangle+\langle K-|V|K-\rangle,
 \nonumber\\
 2v_z
 &=\langle K+|V|K+\rangle-\langle K-|V|K-\rangle,
 \nonumber\\
 v_x-iv_y
 &=\langle K+|V|K-\rangle .
 \label{eq:effective-K-components}
\end{align}
The terms $v_0I_2+v_z\tau_z$ are $K$ preserving, whereas $v_x\tau_x+v_y\tau_y$ arise from the projected $K$-odd interaction. Defining
\begin{equation}
 v_\perp^2=v_x^2+v_y^2
 =|\langle K+|V|K-\rangle|^2,
 \qquad
 \Omega_K=\sqrt{v_\perp^2+v_z^2},
 \label{eq:K-Rabi-frequency}
\end{equation}
a constant pulse of duration $t$ gives, for an initial $|K+\rangle$ state,
\begin{equation}
 P_{+\to-}(t)
 =\frac{v_\perp^2}{v_\perp^2+v_z^2}
 \sin^2\!\left(\Omega_K t\right).
 \label{eq:exact-K-conversion}
\end{equation}
Equation~\eqref{eq:exact-K-conversion} gives the finite-time realization of the symmetry constraints. For a $K$-preserving perturbation, $v_\perp=0$ and $P_{+\to-}=0$ at all times. A transverse gradient may generate $v_\perp\neq0$, whereas a diagonal term $v_z$ reduces the maximum conversion. At short times,
\begin{equation}
 P_{+\to-}(t)=v_\perp^2t^2+O(t^4),
 \label{eq:short-time-K-conversion}
\end{equation}
which is consistent with the quadratic weak-gradient scaling inferred above. The reduction in Eq.~\eqref{eq:effective-K-Hamiltonian} is an effective two-mode description; if the perturbation strongly changes the radial or longitudinal quantum numbers, the full multimode problem must be retained.

\subsection{State preparation, branch analysis, and experimental considerations}
\label{subsec:preparation-readout}

The Hadamard-form relation in Eq.~\eqref{eq:K-Hadamard-basis} connects the $K$ branches to the conventional cylindrical pair. In matrix form,
\begin{equation}
 \begin{pmatrix}
 |K+\rangle\\|K-\rangle
 \end{pmatrix}
 =\mathsf H
 \begin{pmatrix}
 |\uparrow\rangle\\|\downarrow\rangle
 \end{pmatrix},
 \qquad
 \mathsf H=\frac{1}{\sqrt2}
 \begin{pmatrix}
 1&1\\1&-1
 \end{pmatrix},
 \label{eq:Hadamard-matrix-operational}
\end{equation}
with $\mathsf H^{-1}=\mathsf H$. A coherent superposition of the conventional pair with relative phase $0$ or $\pi$ prepares the two $K$ branches, respectively, and the inverse transformation maps branch amplitudes onto conventional-mode populations.

For a general evolution matrix $S_K$ in the branch basis, the corresponding conventional-basis matrix is
\begin{equation}
 S_{\rm conv}=\mathsf H S_K\mathsf H.
 \label{eq:general-Hadamard-readout}
\end{equation}
When the evolution preserves $K$ in the selected two-mode sector,
\begin{equation}
 S_K=
 \begin{pmatrix}
 s_+&0\\
 0&s_-
 \end{pmatrix},
 \label{eq:SK-diagonal}
\end{equation}
Eq.~\eqref{eq:general-Hadamard-readout} gives
\begin{equation}
 S_{\rm conv}
 =\frac12
 \begin{pmatrix}
 s_++s_-&s_+-s_-\\
 s_+-s_-&s_++s_-
 \end{pmatrix}.
 \label{eq:S-conventional}
\end{equation}
For an incident $|\uparrow\rangle$ mode,
\begin{align}
 \mathcal A_{\uparrow\to\uparrow}
 &=\frac{s_++s_-}{2},
 &
 \mathcal A_{\uparrow\to\downarrow}
 &=\frac{s_+-s_-}{2}.
 \label{eq:conventional-amplitudes}
\end{align}
A $K$-preserving interaction need not distinguish the two branches. If $s_+=s_-$, it produces a common amplitude and no population contrast in this analysis. When $s_+\neq s_-$, however, their relative amplitude or phase is converted into a population contrast in the conventional basis. In the phase-only case,
\begin{equation}
 s_\eta=e^{i\phi_\eta},
\end{equation}
one obtains
\begin{equation}
 P_{\uparrow\to\downarrow}
 =\sin^2\!\left(\frac{\Delta\phi}{2}\right),
 \qquad
 \Delta\phi=\phi_+-\phi_- .
 \label{eq:phase-readout}
\end{equation}
Branch populations and branch-relative phases therefore require distinct analyses. A branch-resolved measurement probes conversion, whereas the Hadamard-form analysis maps a diagonal branch difference onto population transfer in the conventional basis.

The preceding results suggest three distinct classes of measurements. Propagation of a coherently prepared branch through a transversely uniform longitudinal potential $V(z,t)$ tests the predicted vanishing of the opposite-branch amplitude in Eq.~\eqref{eq:null-branch-flip}. A continuously tunable radial component of the form in Eq.~\eqref{eq:weak-radial-profile} permits the onset of $\sigma_{\rm flip}$, or of the oscillations in Eq.~\eqref{eq:exact-K-conversion}, to be compared with the projected $K$-odd matrix element. Finally, if a $K$-preserving interaction produces unequal diagonal amplitudes, the inverse Hadamard transformation permits their relative phase or attenuation to be inferred from conventional-mode populations.

Electron-optical mode converters and orbital-angular-momentum analyzers demonstrate coherent control and analysis of structured electron modes~\cite{Grillo2017OAMSorter,Tavabi2021}. Such devices do not by themselves measure $K$, because a $K$ eigenmode correlates orbital harmonics with spinor components. An implementation would require coherent preparation of the two conventional cylindrical modes, control of their relative phase, and an inverse Hadamard-form analysis combined with spin-sensitive mode detection. The conversion efficiency remains dependent on the finite beam profile and the spatial structure of the interaction region.

\section{Conclusion}
\label{sec:conclusion}

We have constructed the general regular positive-energy cylindrical solution of the free Dirac equation at fixed $(E,\kappa,k_z,J_z)$. The resulting two-dimensional solution space provides a common framework for mode constructions that are usually introduced separately. The conventional spin-polarized vortex modes and the Mihaila--Dawson--Cooper separation modes occur as special members of this family, while helicity-adapted twisted states follow from a unitary change of basis. The different formulations therefore represent distinct internal organizations of the same regular cylindrical sector.

The remaining twofold freedom is resolved by the Hermitian transverse operator $K=\beta(\boldsymbol{\Sigma}\times\mathbf p)_z$, which commutes with the free Hamiltonian, the longitudinal momentum, and the total angular momentum. Its eigenstates reproduce, up to conventions, the transverse-helicity cylindrical basis known from earlier quantization and subsequent rotating and confined Dirac analyses~\cite{BalantekinDeWeerd1995,JiangLiao2016,Khosravi2019PRB}, while the present construction derives this branch resolution directly from the general regular free family and relates it explicitly to the conventional spin, helicity, and separation bases. This symmetry resolution also supplies the natural setting for the quantum-geometric analysis. The unresolved positive-energy sector is rank two and admits momentum-dependent changes of basis, whereas $K$ selects two one-dimensional eigenspaces and hence two gauge-invariant joint spectral projectors. Their geometry is therefore symmetry defined rather than attached to an arbitrary basis choice: the two branches carry opposite Berry curvatures, possess the same quantum metric, and saturate the metric--curvature relation of a two-level system. The construction characterizes the internal positive-energy sector and is distinct from geometric contributions associated with a finite spatial envelope.

The same symmetry organizes the branch-resolved dynamics. A general perturbation separates into $K$-even and $K$-odd components, which respectively preserve and mix the branches. For scalar potentials, a transversely uniform longitudinal profile conserves $K$, whereas a radial gradient produces branch-off-diagonal coupling. Projection onto an isolated positive-energy doublet gives a closed two-state evolution without assigning an artificial splitting to the degenerate free branches. In addition, the Hadamard-form relation to the conventional cylindrical pair converts branch-dependent amplitudes and phases into conventional-mode population contrasts, subject to the mode-selectivity and detection requirements discussed in Sec.~VI.

These results establish a unified exact solution family together with a symmetry-adapted branch classification, its intrinsic Abelian geometry, and the corresponding dynamical selection rules. More broadly, they show that cylindrical Dirac modes appearing in physically distinct axially symmetric problems can be understood as different resolutions of a common relativistic single-particle sector. A gauge-covariant treatment of the full degenerate positive-energy sector, including its non-Abelian structure and alternative symmetry reductions, is left for subsequent work.

\begin{acknowledgments}
Q.G. acknowledges financial support from the National Natural Science Foundation of China through Grant No.~11874083.
\end{acknowledgments}

	\bibliography{references}

\end{document}